\documentclass[twocolumn,pre]{revtex4}
\usepackage{amscd}
\usepackage{amssymb}
\usepackage{amsfonts}
\usepackage{amsmath}
\usepackage{graphicx}
\usepackage{dcolumn}
\usepackage{bm}
\usepackage[dvipdfm]{hyperref}

\begin{document}

\preprint{}
\title{Revising the simple measures of assortativity in complex networks}
\author{Xiao-Ke Xu $^{1,2}$}
\email{xiaokeeie@gmail.com}
\author{Jie Zhang $^2$}
\author{Junfeng Sun $^3$}
\author{Michael Small $^2$}
\affiliation{$^1$School of Communication and Electronics
Engineering,
Qingdao Technological University, Qingdao 266520, China\\
$^2$Department of Electronic and Information
Engineering, Hong Kong Polytechnic University, Hung Hom, Kowloon, Hong Kong\\
$^3$Med-X Research Institute, Shanghai Jiao Tong University, 1954
Hua Shan Road, Shanghai 200030, China}
\date{\today}

\begin{abstract}
We find that traditional statistics for measuring degree mixing are
strongly affected by superrich nodes. To counteract and measure the
effect of superrich nodes, we propose a paradigm to quantify the
mixing pattern of a real network in which different mixing patterns
may appear among low-degree nodes and among high-degree nodes. The
new paradigm and the simple revised measure uncover the true complex
degree mixing patterns of complex networks with superrich nodes. The
new method indicates that some networks show a false disassortative
mixing induced by superrich nodes, and have no tendency to be
genuinely disassortative. Our results also show that the previously
observed fragility of scale-free networks is actually greatly
exacerbated by the presence of even a very small number of superrich
nodes.
\end{abstract}

\maketitle

\section{Introduction}
With increasing evidence of the ubiquity of scale-free networks,
attention has recently shifted to the particular structure of
experimentally observed networks \cite{Correlation_Science,
Correlation_Scale, Review_Newman, Phycics_today,
Extreme_assortativity1, Extreme_assortativity2}. One key observation
to arise has been that some networks display a propensity for high
degree nodes to connect to other high degree nodes ({\em
assortativity}) \cite{Mixing_Newman02, Social_mixing}. Conversely in
certain other types of networks the reverse is true: high degree
nodes connect to low degree nodes ({\em disassortativity}). In
particular, numerical evidence from experimental data has shown that
many technological (i.e. communication networks), biological (e.g.,
protein and neural networks) and certain social (online communities
\cite{Nioki_SN, Online_SN}) networks are found to exhibit a negative
assortativity coefficient and are therefore claimed to be examples
of disassortative mixing \cite{Mixing_Newman02, Mixing_Newman03}.

A widely accepted way to determine the mixing pattern of complex
networks is to calculate the Pearson correlation coefficient of the
degrees at both ends of the edges \cite{Review_Newman,
Review_measure, Review_Boccaletti}. Although Newman's assortativity
coefficient is not a perfect measure of assortativity, it is a very
simple measure to approximately assess the degree mixing pattern of
a network and has been employed broadly. Another natural approach to
quantify mixing patterns is to calculate the correlations between
two nodes connected by an edge \cite{Review_measure}. There are two
ways to characterize the node degree correlations. One is the
conditional probability $P(k'|k)$ that an arbitrary edge connects a
node with degree $k$ to another node with degree $k'$ \cite{Kn}. The
other is the joint degree distribution $P(k,k')$ that measures
whether nodes with a given degree $k$ prefer (or avoid) to connect
to nodes with degree $k'$ \cite{Correlation_Science,
Correlation_Internet, Correlation_Scale}. Although they can capture
the mixing patterns by the 2-dimensional degree correlation plots
\cite{Internet_mixing, Knn1, Knn2}, it is difficult for the two
methods to quantitatively evaluate the assortativities of complex
networks.

While the existing measures for assortativity are simple, and the
mathematical definition of what is meant by assortativity and
disassortativity is clear and consistent, this is not equivalent to
the general understanding of these phenomena. In particular,
disassortativity can easily and often does arise in situations when
the degree of neighbors matches as closely as possible. This is
completely at odds with what is commonly understood, and this needs
to be addressed. The current usage of degree correlation measures
means that researchers often conclude that node degree between
neighbors in physical/social systems is mis-matched while in fact
the opposite is true.

In this paper we show that in a finite size network superrich nodes
cause a network to show an observed disassortative mixing in most
cases, while they sometimes let a strong disassortative network
appear less disassortative property. And superrich nodes often limit
the assortativity coefficient to be in a narrow range, and mask the
genuine mixing pattern of complex networks. The superrich nodes,
which exist in many real networks, refer to the nodes whose degrees
are far larger than most other nodes. The assortativity coefficient
in certain experimentally measured networks is therefore due to a
small fraction of superrich nodes, and is not caused by a genuinely
assortative (or disassortative) mixing. The mechanism underlying
degree mixing has not been properly understood and many of these
common indices mis-represent the true assortativity of a network. We
find that the assortativity coefficient will give a ``false" result
of mixing patterns for the networks with superrich nodes.

To overcome the effect of these aberrant and uncommonly connected
nodes, a new and more robust measure of degree-degree mixing is
needed. We choose a modification of Newman's assortativity
coefficient so that our new measure will be simple and also as
similar as possibly to what is already being employed. We
deliberately choose to change the existing measure as little as
possible, so that the new measure can be best understood. We are not
suggesting that the existing tools be abandoned and only that they
be computed twice: once as is done now, and again after removing the
contribution of superrich nodes.

The superrich nodes can, and often will, have an extremely great
effect on the network structure. Our results also show that the
previously observed fragility of scale-free networks is actually
greatly exacerbated by the presence of even a very small number of
superrich nodes. An attack targeting the \emph{superrich} nodes
(rather than just the \emph{rich}) of a network can very quickly
fragment, or at least stretch, a scale-free network.

\section{Superrich nodes affect degree mixing patterns}
\subsection{Superrich nodes in scale-free network}
The superrich nodes refer to the nodes whose degrees are far larger
than most other nodes. Our finding in this study can be applied to
any degree distribution network, and is not limited for power-law
and exponential degree distribution networks, so there is no need to
judge whether the degree distribution obeys a power-law strictly.
While it is necessary to supply a simple and operative definition of
superrich nodes in scale-free networks, for the power-law degree
distribution broadly exists in real networks \cite{BA}. In terms of
networks with approximately power-law (or exponential) degree
distribution, \emph{superrich} nodes are defined as the nodes whose
degrees are larger than the natural cutoff value. It should be
noticed that the nodes with degree that is predicted by the
power-law only are \emph{rich} nodes.

The natural cutoff degree $k_c$ is an important concept in
finite-size scale-free networks \cite{Degree_cutoff}. It is defined
as the value of the degree $k_c$ above which one expects to find at
most one node \cite{Cutoff_value}, that is
\begin{equation}
 N \int_{k_c}^\infty P(k) d k \sim 1,
\end{equation}
\noindent here $N$ is the number of nodes. For a scale-free network,
this expression provides a dependence of the natural cutoff with $N$
and the slope $\gamma$ as
\begin{equation}
k_c(N) \sim N^{1/(\gamma-1)}.
\end{equation}
Here we obtain the slope $\gamma$ by fitting the real data excluding
the potential superrich nodes.

\subsection{Adding superrich nodes to BA model}
\begin{figure}[htbp]
\centering
\includegraphics[width=0.44\textwidth]{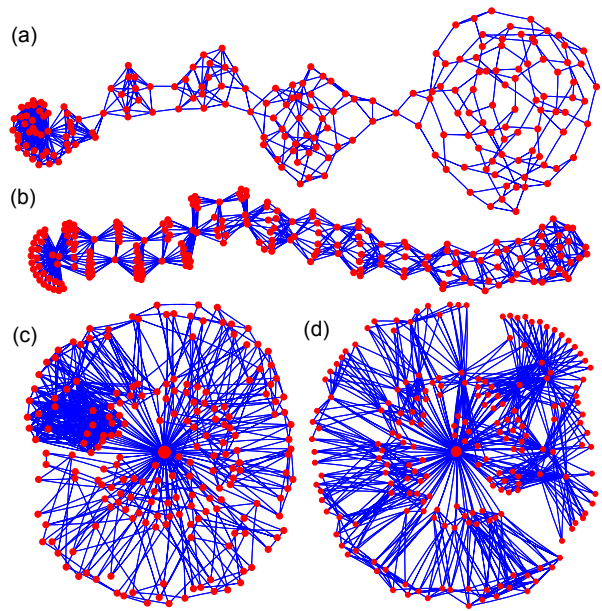}
\caption{(Color online) Networks without and with a superrich node
show greatly different topological structure, average path length
$l$ and clustering coefficient $c$: (a) maximal assortative mixing
network, $l=9.73$ and $c=0.23$; (b) maximal disassortative mixing
network, $l=8.19$ and $c=0.02$; (c) maximal assortative mixing
network with one superrich node ($k=100$) added randomly, $l=2.88$
and $c=0.30$; (d) maximal disassortative mixing network with one
superrich node ($k=100$) added randomly, $l=2.75$ and $c=0.10$.}
\label{BA_Origin_Mixing}
\end{figure}

First we utilize the BA model \cite{BA} to demonstrate the
significant effect of superrich nodes. We generate a small size of
BA network (nodes $n=200$ and average degree $\left<k\right>=6$),
and then the maximal assortative mixing (MAM) and maximal
disassortative mixing (MDM) networks are generated from the original
BA network using the rewiring method \cite{Constraint}. These three
networks have the same degree distribution but different mixing
patterns. The MAM network reveals a beautiful assortative structure
in which nodes with similar degrees connect to one another as is
shown in Fig. \ref{BA_Origin_Mixing}(a). In contrast, the MDM
network shows that high degree nodes link to low degree ones in Fig.
\ref{BA_Origin_Mixing}(b). However, the addition of only one
superrich node ($k=100$ and random linking) severely affects the
entire topology of the network, as is shown in Fig.
\ref{BA_Origin_Mixing}(c) and (d). Our results in Fig.
\ref{BA_Origin_Mixing} also show that the clustering coefficient
\cite{WS} and average path length are extremely changed by the only
one superrich node.

Now we demonstrate how superrich nodes bias the mixing pattern of a
network. The assortativity coefficient $r$ \cite{Mixing_Newman02} is
given by

\begin{equation}
r = {M^{-1} \sum_i j_i k_i -
    \bigl[ M^{-1} \sum_i \mbox{$\frac12$}(j_i+k_i) \bigr]^2\over
    M^{-1} \sum_i \mbox{$\frac12$}(j_i^2+k_i^2) -
    \bigl[ M^{-1} \sum_i \mbox{$\frac12$}(j_i+k_i) \bigr]^2},
\end{equation}

\noindent where $j_i$ and $k_i$ are the degrees of the two endpoints
of the $i$th edge, and $M$ is the total number of edges in the
network. If $r>0$, the network is claimed to be assortative mixing;
while if $r<0$, the network is called disassortative mixing.

A large scale BA network ($n=5000$ and $\left<k\right>=6$) is
generated, and the MAM and MDM networks are obtained by the rewiring
method \cite{Constraint}. The $r$ values for these networks are
listed in Table \ref{BA_table}. The above three networks exhibit the
reasonable mixing coefficients: assortative, neutral and
disassortative. Then we add five superrich nodes ($k=1000$) to each
of the three networks. For the MAM network, superrich nodes are
connected to high degree nodes; while for the MDM network, superrich
nodes are linked to low degree nodes; for the original BA network,
random linking is adopted. As can be seen in Table \ref{BA_table},
all $r$ are negative for these networks, which runs against our
intuition. Take the MAM network for example, the strong positive $r$
($0.320$) is replaced by a negative value ($-0.136$), although the
superrich nodes are attached to the high degree nodes (assortative
adding). Moreover, the fluctuation of $r$ becomes narrow for the
networks with superrich nodes ($[-0.189, 0.320]\Rightarrow[-0.169,
-0.136]$). This result shows that superrich nodes can not only make
a strong assortative network show a negative assortativity
coefficient, but also make a strong disassortative network appear
less disassortative.

\begin{table}[!h]
\tabcolsep 0pt \caption{Assortativity coefficients $r$ of different
mixing patterns for BA networks without and with superrich nodes.}
\label{BA_table}

\vspace*{-12pt}
\begin{center}
\def\temptablewidth{0.46\textwidth}
{\rule{\temptablewidth}{1pt}}
\begin{tabular*}{\temptablewidth}{@{\extracolsep{\fill}}l c}
Network & $r$ \\
\hline
Maximal disassortative mixing  & $-0.189\pm{0.029}$ \\

Original BA network  & $-0.047\pm{0.004}$ \\

Maximal assortative mixing  & $0.320\pm{0.088}$ \\
\hline
Maximal disassortative mixing with $5$ & \\
superrich nodes (disassortative adding) & \raisebox{1.3ex}[0pt]{$ -0.169\pm{0.002}$} \\
\hline
Original BA network with $5$ superrich& \\
nodes (random adding) & \raisebox{1.3ex}[0pt]{$ -0.163\pm{0.002}$} \\
\hline
Maximal assortative mixing with $5$ & \\
superrich nodes (assortative adding) &
\raisebox{1.3ex}[0pt]{$-0.136\pm{0.002}$}
\end{tabular*}
{\rule{\temptablewidth}{1pt}}
\end{center}
\end{table}

\subsection{Superrich nodes in experimental networks}
\begin{figure}[htbp]
\centering
\includegraphics[width=0.45\textwidth]{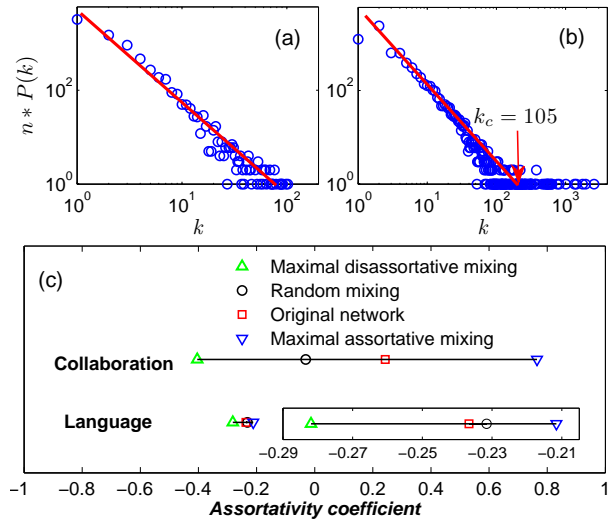}
\caption{(Color online) Degree distributions of two real networks
and their assortativity coefficients for different mixing patterns:
(a) authors collaboration network in computational geometry (nodes
$n=7343$, edges $m=11898$ and maximum degree $k_{max}=102$)
\cite{Pajek_data}; (b) language network, which is a word adjacency
network of texts from Darwin's ``The Origin of Species'' ($n=7724$,
$m=46281$ and $k_{max}=2568$) \cite{Superfamily}; (c) assortativity
coefficients of different mixing patterns for the above two
networks.} \label{Degree_Distribution}
\end{figure}

Many complex networks exhibit a scale-free degree distribution
\cite{BA, Review_Barabasi}, such as the two real networks in Figs.
\ref{Degree_Distribution}(a) and (b). And the slope of power-law is
obtained by fitting the real data excluding the potential superrich
nodes. For the collaboration network in computational geometry
\cite{Pajek_data}, its degree fits the power-law distribution
perfectly. While for the language network \cite{Superfamily}, from
Darwin's ``The Origin of Species'', superrich nodes exist in it,
because the maximum degree ($k_{max}=2568$) is far larger than the
natural cutoff ($k_c=105$). The results of $r$ for four mixing
patterns of the two networks are shown in Fig.
\ref{Degree_Distribution}(c). The original collaboration network is
assortative like most social networks \cite{Mixing_Newman02,
Social_mixing}. While for all mixing patterns of the language
network, $r$ is strong negative and is confined to a narrow range,
which is similar to the BA network with superrich nodes. According
to the results of the theoretical model (BA network) and the
language network, we conclude that $r$ can not determine mixing
patterns of networks with superrich nodes accurately, and superrich
nodes lead $r$ to be strong negative and within a very narrow range.

One method to characterize the node degree correlations is the
conditional probability $P(k'|k)$ that an arbitrary edge connects a
node with degree $k$ to another node with degree $k'$ \cite{Kn}. It
is difficult for this method to quantitatively evaluate mixing
patterns of complex networks with superrich nodes, because the
finite network size and the small amount of superrich nodes will
lead to an unstable result \cite{Review_measure, Constraint}. This
problem can be partially solved by calculating the average degree of
the nearest neighbors of nodes with a given degree $k$ \cite{Knn1,
Knn2}, which is given by

\begin{equation}
\left<K_{nn}\right> = \sum_{k'} k' P(k'|k). \label{knn}
\end{equation}
Here an increasing $\left<K_{nn}\right>$ with $k$ indicates that
nodes with high degree tend to connect to nodes with high degree,
and the network is classified as assortative. Whereas a decreasing
$\left<K_{nn}\right>$ with $k$ indicates that nodes with high degree
tend to connect to nodes with low degree, and the network is
disassortative.

\begin{figure}[htbp]
\centering
\includegraphics[width=0.45\textwidth]{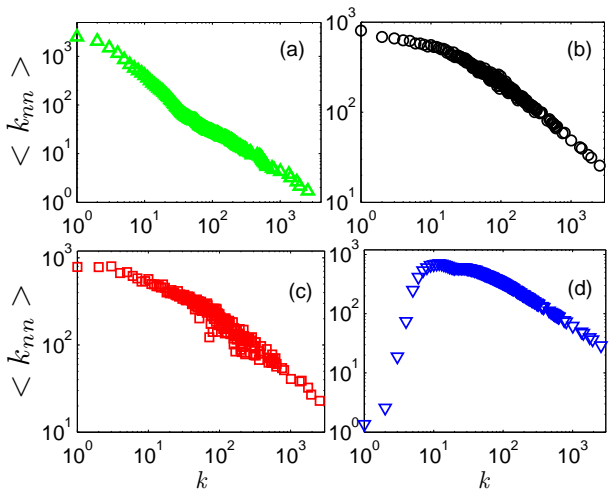}
\caption{(Color online) Average degree $\left<K_{nn}\right>$ of the
nearest neighbors of a node depending on its degree $k$ for
different mixing patterns of the language network: (a) maximal
disassortative mixing, (b) random mixing, (c) original network and
(d) maximal assortative mixing.} \label{Knn}
\end{figure}

The average degree $\left<K_{nn}\right>$ for different mixing
patterns of the language network is shown in Fig. \ref{Knn}. For the
MDM, original and random mixing networks, $\left<K_{nn}\right>$ all
decrease with $k$, which shows a disassortative property. For the MAM
network, $\left<K_{nn}\right>$ firstly increases and then decreases,
and thus it is difficult to tell whether the network is assortative
or disassortative. These results indicate that neither $r$ nor
$\left<K_{nn}\right>$ can characterize the intrinsic mixing patterns
of complex networks with superrich nodes, for the degree of a
superrich node is far larger than those of other nodes.

\section{New paradigm of measuring mixing patterns}
\subsection{New paradigm}
\begin{figure}[htbp]
\centering
\includegraphics[width=0.45\textwidth]{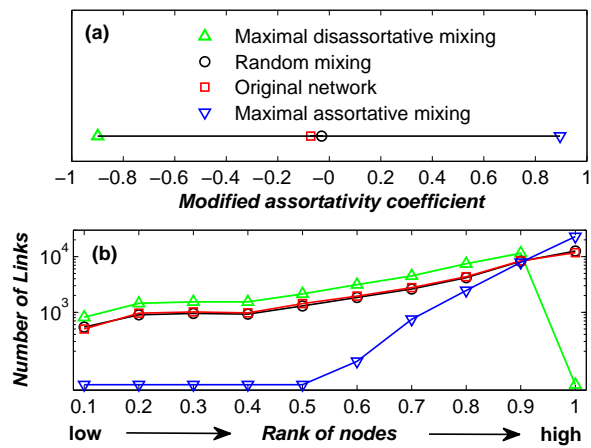}
\caption{(Color online) Results of two new measures for different
mixing patterns of the language network: (a) modified assortativity
coefficient to quantitatively measure mixing patterns among the
nodes whose degrees are below natural cutoff degree; (b) qualitative
measure for the links between superrich nodes and other nodes
(including links among superrich nodes).} \label{new_AS}
\end{figure}

Since superrich nodes are widely observed in real networks, a new
statistic is needed to appropriately classify whether a network with
superrich nodes is genuinely assortative or not. In this paper, we
manage to determine the mixing patterns of complex networks with two
new measures. The first one is a modified definition of
assortativity coefficient, which is given by

\begin{equation}\label{Rc}
r_c = {M_c^{-1} \sum_i j_i k_i -
    \bigl[ M_c^{-1} \sum_i \mbox{$\frac12$}(j_i+k_i) \bigr]^2\over
    M_c^{-1} \sum_i \mbox{$\frac12$}(j_i^2+k_i^2) -
    \bigl[ M_c^{-1} \sum_i \mbox{$\frac12$}(j_i+k_i) \bigr]^2},
\end{equation}

\noindent where $M_c$, which is different to the original $M$, is
the total number of edges among the nodes whose degrees are lower
than $k_c$. In accordance with $r$, positive $r_c$ indicates a
network is assortative; and negative $r_c$ represents a network is
disassortative. When calculating $r_c$ in equation \ref{Rc}, we
exclude superrich nodes whose degrees are greater than $k_c$. In
contrast with the failure of $r$, $r_c$ can determine the genuine
mixing patterns of complex networks with superrich nodes and
distinguish the four mixing patterns very effectively as is
demonstrated in Fig. \ref{new_AS}(a). Especially, $r_c$ of the
original language network is neutral, which is different to the
strong negative $r$ that we have observed in Fig.
\ref{Degree_Distribution}(c).

Although our new statistic $r_c$ can effectively measure the mixing
patterns among the nodes whose degree is below the natural cutoff
value, it is not an optimal solution to neglect a very small number
but the most important superrich nodes. Moreover, a real network may
exhibit assortative mixing among low degree nodes and disassortative
mixing among superrich nodes, and it is not accurate to assert the
network is assortative or disassortative. It is necessary to propose
a paradigm to measure the mixing pattern of low-degree nodes and
high-degree nodes respectively, especially for the network with
superrich nodes, because the assortativity coefficient of a very
small number of superrich nodes will mask the mixing pattern of
larger number of low degree nodes.

We have used the modified assortativity coefficient to
quantitatively measure the mixing pattern of low degree nodes, so
the second measure is proposed to quantify the mixing pattern
between low degree nodes and superrich nodes (including links among
superrich nodes). Because the degrees of superrich nodes are far
larger than most other nodes and superrich nodes have to link to
many low degree nodes, it is meaningless to say the mixing pattern
of superrich nodes is always disassortative. To find the genuine
mixing pattern of superrich nodes, here we test superrich nodes tend
to link to high or low \emph{rank} nodes. If superrich nodes tend to
link to high rank nodes, we consider that they are assortative
mixing. Conversely, if superrich nodes tend to link to low rank
nodes, we believe that they show a disassortative property.

Firstly nodes are ranked in an increasing order according to the
degrees. If two nodes have the same degree, we rank their orders
randomly. The nodes are divided into $10$ bins according to their
ranks. Then the number of links connecting superrich nodes to the
nodes in each bin is calculated. We can not calculate the
assortativity coefficient of the orders after ordering all nodes
because of a false result given by the large number of links of
superrich nodes. For different mixing patterns, the distributions of
linking numbers are shown in Fig. \ref{new_AS}(b). The MAM network
shows assortative for superrich nodes tend to connect to \emph{high
rank} nodes. On the contrary, the MDM network shows disassortative
for superrich nodes tend to link to \emph{low rank} nodes. The
original network and the random mixing network are both neutral
mixing for superrich nodes have no obvious tendency to connect with
high or low rank nodes.

Figures \ref{new_AS}(a) and (b) show the coincident results for
different mixing patterns, which indicates the two new measures are
both effective. Furthermore, our two new measures compose an
effective paradigm to quantify a real network in which different
mixing patterns may appear among low-degree nodes and among
high-degree nodes. They quantify the effect of superrich nodes on
the mixing pattern of complex networks. The new paradigm can capture
the rich nature of the mixing properties in a real network and
distinguish whether the disassortativity of a network is derived
from superrich nodes.

\subsection{Comparison with correlation profile}
\begin{figure*}[htbp]
\centering
\includegraphics[width=1.0\textwidth]{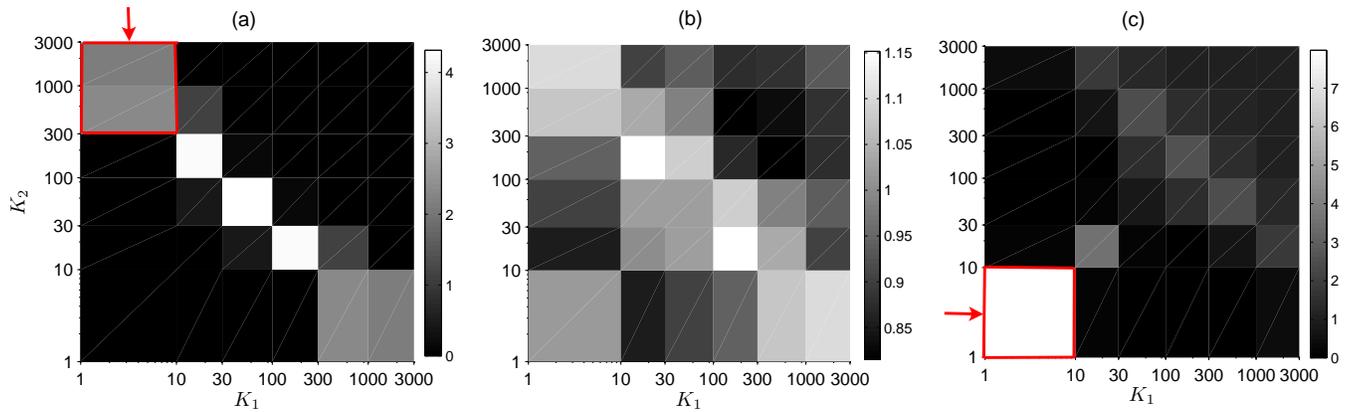}
\caption{(Color online) Correlation profile for different mixing
patterns of the language network: (a) maximal disassortative mixing,
(b) original network and (c) maximal assortative mixing. These
figures show the ratio $R(k_1,k_2)=N(k_1,k_2)/N_r(k_1,k_2)$, where
$N(k_1,k_2)$ is the total number of edges in the links of the
language network with degrees $k_1$ and $k_2$, while $N_r(k_1,k_2)$
is the same value in a randomized version of this network.}
\label{Correlation_profile}
\end{figure*}

The joint degree distribution $P(k,k')$ is another way to
characterize the node degree correlations, which can measure whether
nodes with a given degree $k$ prefer (or avoid) to connect to nodes
with degree $k'$. The 2-dimensional degree correlation profile
\cite{Correlation_Science, Correlation_Internet} is a suitable way
to use the joint degree distribution to capture the mixing patterns.
We visualize the correlation profile for different mixing patterns
of the language network by plotting the ratio
$R(k_1,k_2)=N(k_1,k_2)/N_r(k_1,k_2)$ in Fig.
\ref{Correlation_profile}, where $N(k_1,k_2)$ is the total number of
edges in the links of the language network with degrees $k_1$ and
$k_2$, while $N_r(k_1,k_2)$ is the same value in a randomized
version of this network.

Observing the results in Fig. \ref{Correlation_profile}, we find the
correlation profile effectively captures the degree correlations of
neighboring nodes. The left-up corner of Fig.
\ref{Correlation_profile}(a) shows that high degree nodes tend to
link to low degree nodes (disassortative) in the MDM network.
Conversely, the MAM network obviously shows that low degree nodes
prefer to connect to low degree nodes (assortative) in the left-down
corner of Fig. \ref{Correlation_profile}(c). All the values of
$R(k_1,k_2)$ for the original language network are in a very narrow
range $[0.82, 1.15]$ in Fig. \ref{Correlation_profile}(b), which
means this network is near neutral mixing. These results are
consistent with those in Figs. \ref{new_AS}(a) and (b), which shows
that the 2-dimensional degree correlation profile is an effective
way to capture the mixing patterns and it is more robust than
original $r$ and $\left<K_{nn}\right>$ to capture the mixing
patterns of complex networks with superrich nodes.

Comparing the correlation profile with our new paradigm, we can find
the advantage of 2-dimensional correlation profile is describing the
detailed information of degree correlations. But the correlation
profile is only a qualitative method and is difficult to
quantitatively evaluate the assortativities of complex networks. The
assortativity coefficient is a very simple and the most widely
accepted quantitative way to determine the degree mixing pattern. We
choose a modification of Newman's assortativity coefficient $r_c$ as
the measure for real networks so that our new measure will be simple
and also as similar as possible to what is already being employed.
And this measure avoids replacing the existing simple and useful
measures with something entirely different.

\section{Results of experimental networks}
The results of $r$ and $r_c$ for nine real networks are listed in
Table \ref{Real_results}. Like most social networks, networks (a)
and (b) have positive $r$, for the nodes tend to connect to one
another belonging to the same community \cite{Social_mixing} and
there are no superrich nodes. But $r$ of Wealink.com is negative,
which is consistent with other online communities \cite{Nioki_SN,
Online_SN}. And $k_c$ of Wealink.com is $150$, so a person who has
more than $150$ links is a \emph{superstar}. Although the number of
superstars in Wealink.com is only $166$ ($0.07\%$), surprisingly
$r_c$ becomes more negative ($-0.08 \Rightarrow -0.42$) after
excluding these superstars. The above result indicates that a very
small number of superrich nodes mask the genuine degree mixing
information of the online community. And superrich nodes do not
always make a strong assortative network show a negative
assortativity coefficient. On the contrary, sometimes they will make
a strong disassortative network appear less disassortative property,
just like in this case.

\begin{table}[htbp]
\tabcolsep 0pt \caption{Properties of real undirected networks:
number of nodes $n$, edges $m$, natural cutoff degree $k_{c}$ ,
maximal degree $k_{max}$, original assortativity coefficient $r$,
and modified assortativity coefficient $r_c$. Note that for the
networks without superrich nodes, $k_c=k_{max}$ and $r_c=r$, and
thus the values of $k_c$ and $r_c$ are not listed. Social networks:
(a) network of e-mail interchanges between members in a university
\cite{Alex_datasets}; (b) collaboration network of scientists who
work on the condensed matter \cite{Newman_datasets}; (c) online
social network of Wealink.com \cite{Ruolin}. Technological networks:
(d) network of P.M. Roget's Thesaurus \cite{Pajek_data}; (e) network
of articles by and citing J. Lederberg from 1945 to 2002
\cite{Pajek_data}; (f) Internet snapshot at the level of autonomous
systems \cite{Newman_datasets}. Biological networks: (g) network of
metabolic pathways for \emph{E. Coli} \cite{Cosin_datasets}; (h)
network of protein folding \cite{Cosin_datasets}; (i) neural network
of \emph{C. Elegans} \cite{WS}.}

\vspace*{-12pt}
\begin{center}
\def\temptablewidth{0.48\textwidth}
{\rule{\temptablewidth}{1pt}}
\begin{tabular*}{\temptablewidth}{@{\extracolsep{\fill}}l c c c c c c}
Network & $n$ & $m$ & $k_{c}$ & $k_{max}$ & $r$ & $r_c$\\
\hline
(a) e-mail & $1134$ & $10902$ & $-$ & $71$ & $0.08$ & $-$ \\
(b) arXiv.org & $40421$ & $175693$ & $-$ & $278$ & $0.19$ & $-$ \\
(c) Wealink.com & $223624$ & $273395$ & $150$ & $1657$ & $-0.08$ & $-0.42$ \\

\hline
(d) Roget's thesaurus & $1022$ & $5103$ & $-$ & $28$ & $0.18$ & $-$ \\
(e) citation network & $8843$ & $41609$ & $105$ & $1103$ & $-0.10$ & $-0.02$ \\
(f) Internet & $22963$ & $48436$ & $110$ & $2390$ & $-0.20$ & $-0.29$ \\

\hline
(g) metabolic pathway & $896$ & $964$ & $-$ & $18$ & $0.15$ & $-$ \\
(h) protein folding & $1287$ & $33813$ & $-$ & $319$ & $0.17$ & $-$ \\
(i) neural net & $307$ & $2359$ & $38$ & $134$ & $-0.16$ & $0.03$

\end{tabular*}
{\rule{\temptablewidth}{1pt}} \label{Real_results}
\end{center}
\end{table}

Usually it is believed that technological and biological networks
tend to be disassortative \cite{Mixing_Newman02, Mixing_Newman03},
such as the networks (e), (f) and (i). We find that superrich nodes
lead $r$ of these networks to be negative. After filtering superrich
nodes, (e) and (i) show neutral mixing properties, while $r_c$ of
(f) turns more negative. The above results show that $r_c$ can
reveal the intrinsic mixing patterns masked by superrich nodes.
Since previous works do not take the effect of superrich nodes into
consideration, many networks exhibit ``false" negative $r$ and
decreasing $\left<K_{nn}\right>$. The advantage of $r_c$ over the
previous measures is that it is not affected by superrich nodes. The
modified definition of assortativity coefficient $r_c$ is more
suitable to characterize the mixing pattern of any degree
distribution network, especially for a network with superrich nodes.

One possible explanation for the disassortative mixing in these
non-social networks is the structural constraint: two nodes have no
more than one edge connecting them \cite{Correlation_Internet}. But
the structural constraint can not explain why not all non-social
networks are disassortative. Actually, further research has found
the prohibitively multi-edged mechanism can not generate the same
correlation as the real Internet, and only part of degree
correlations can be obtained in this way \cite{Internet_mixing}. And
it is difficult to distinguish which part of disassortativity is
derived from the structural constraint. As far as we know, the
intrinsic mechanism why these non-social networks exhibit
disassortative properties is not entirely clear \cite{Review_Newman,
Phycics_today}.

The effect of superrich nodes on mixing patterns can explain why
some technological and biological networks tend to have a negative
$r$. Superrich nodes exist in these networks more commonly than in
social networks, so many technological and biological networks
exhibit a negative $r$ induced by superrich nodes and are therefore
claimed to be examples of disassortative mixing. Furthermore, we
find that the technological and biological networks (without
superrich nodes) have no tendency to be disassortative, such as the
networks (d), (g) and (h), for a positive $r$ is commonly found in
these networks. Our results indicate the conjecture that the
disassortativity of degree is the normal state of a network in
\cite{Internet_mixing} may not be right. The normal state of degree
mixing pattern in non-social networks is more like the neural mixing
based on our new paradigm.

\section{Conclusion and discussion}
We show that in a finite size network a very small number of
superrich nodes will bias traditional measures of mixing patterns.
We not only report that the traditional measures are not perfect
measures but also propose a revised measure which is better than the
original one and reveals the real pattern of assortativity. Firstly
we propose a straightforward modification to the existing
assortativity coefficient to both measure and counteract the effect
of superrich nodes. Then we develop an effective paradigm to
quantify a real network in which different mixing patterns may
appear among low-degree nodes and among high-degree nodes. Our new
method can capture the rich nature of the mixing properties in a
real network. We can distinguish the disassortativity of a network
derives from the structural constraint, or other reasons like social
and engineering factors.

Our results also indicate that the ``robust yet fragile" nature of
real networks (e.g., Internet) dose not depends on the power-law
degree distribution only, and superrich nodes have the same effect.
For the random failure of the networks, a very small fraction of
nodes with a very large degree will make any degree distribution
network more robust than random graphs. On the other hand, superrich
nodes are the real Achilles' heel of the Internet. The attack
targeting these richest nodes causes the Internet to collapse faster
than the ER graph, even the scale-free model without superrich nodes
(BA network) as is shown in Fig. 3 in \cite{Attack}. Moreover, the
cascade failure of one superrich node can lead more than $20\%$
nodes of the Internet to be disconnected \cite{Cascading_failure}.

We demonstrate that superrich nodes critically change the way in
which complex networks behave, and we have revised the false
disassortative mixing in some non-social networks induced by
superrich nodes. Revealing the intrinsic mixing patterns of complex
networks masked by superrich nodes is crucial to study epidemic
spreading, percolation on complex networks, error and attack
tolerance of real physical systems. Superrich nodes are one of the
principal factors determining many aspects of the behavior of the
overall network, though they are minority in number. We suggest that
greater attention should be payed to the richest nodes when
analyzing finite size network data.

This work is supported by the University Grants Council of the Hong
Kong government (No. PolyU 5268/07E). Junfeng Sun is partly
supported by a grant from the National Natural Science Foundation of
China (No. 60901025).


\end{document}